\documentclass[aps,prl,amsmath,amssymb,twocolumn,superscriptaddress,showpacs]{revtex4-1}
\usepackage{graphicx}				
\usepackage{amssymb}
\usepackage{amsmath}

\usepackage{color}

\begin{document}
\title{Strong Correlations and d+id Superconductivity in Twisted Bilayer Graphene}

\author{Dante M.~Kennes}
\affiliation{Dahlem Center for Complex Quantum Systems and Fachbereich Physik, Freie Universit\"at Berlin, 14195 Berlin, Germany}
\author{Johannes Lischner}
\affiliation{Depts.~of Physics and Materials and the Thomas Young Centre for Theory and Simulation of Materials,
Imperial College London, London, SW7 2AZ, UK}
\author{Christoph Karrasch}
\affiliation{Dahlem Center for Complex Quantum Systems and Fachbereich Physik, Freie Universit\"at Berlin, 14195 Berlin, Germany}

\begin{abstract}
We compute the phase diagram of twisted bilayer graphene near the magic angle where the occurrence of flat bands enhances the effects of electron-electron interactions and thus unleashes strongly-correlated phenomena. Most importantly, we find a crossover between d+id superconductivity and Mott insulating behavior near half-filling of the lowest electron band when the temperature is increased. This is consistent with recent experiments. Our results are obtained using unbiased many-body renormalization group techniques combined with a mean-field analysis of the effective couplings.

\end{abstract}

\maketitle

\textit{Introduction---}
The discovery of correlated-insulator behaviour \cite{cao2018correlated} and unconventional superconductivity (SC) in twisted bilayer graphene (TBG) by Cao et al. \cite{cao2018unconventional} has triggered an intense research effort to understand the phase diagram as well as other physical properties \cite{chung2018,qiao2018} of this system. TBG is a van der Waals material consisting of two graphene layers which are rotated with respect to each other. At certain magic values of the rotation or twist angle, the Fermi velocity at the Dirac points of TBG vanishes resulting in flat bands in the vicinity of the Fermi energy \cite{bistritzer2011moire, dos2012continuum}. For such a system, it is expected that electron-electron interactions play an important role and could potentially lead to the emergence of exotic correlated phases. The unveiling of this type of unconventional superconductivity and Mott physics in TBG is particularly exciting due to its resemblance to the physics of high-$T_c$ superconductors. In fact, the reported ratio \cite{cao2018unconventional} of the superconducting critical temperature to the Fermi temperature -- a hallmark to decide whether superconductors are in the strong or weak coupling limit -- puts experimentally realized TBG near the magic angle in the ballpark of those ratios obtained for  high-$T_c$ cuprates (LSCO,YBCO,BSCCO), iron pnictides or monolayer iron selenide on a STO surface. These reside among the strongest coupling superconductors known today. Thus, TBG provides an intriguing route to study the largely unknown physics of such a superconductor in the extremely controllable framework offered by graphene where the ratio of the interaction to the kinetic energy can be tuned by approaching the magic twist angle and the filling can be modified by a bottom gate.

\begin{figure}[t]
\centering
\includegraphics[width=\columnwidth]{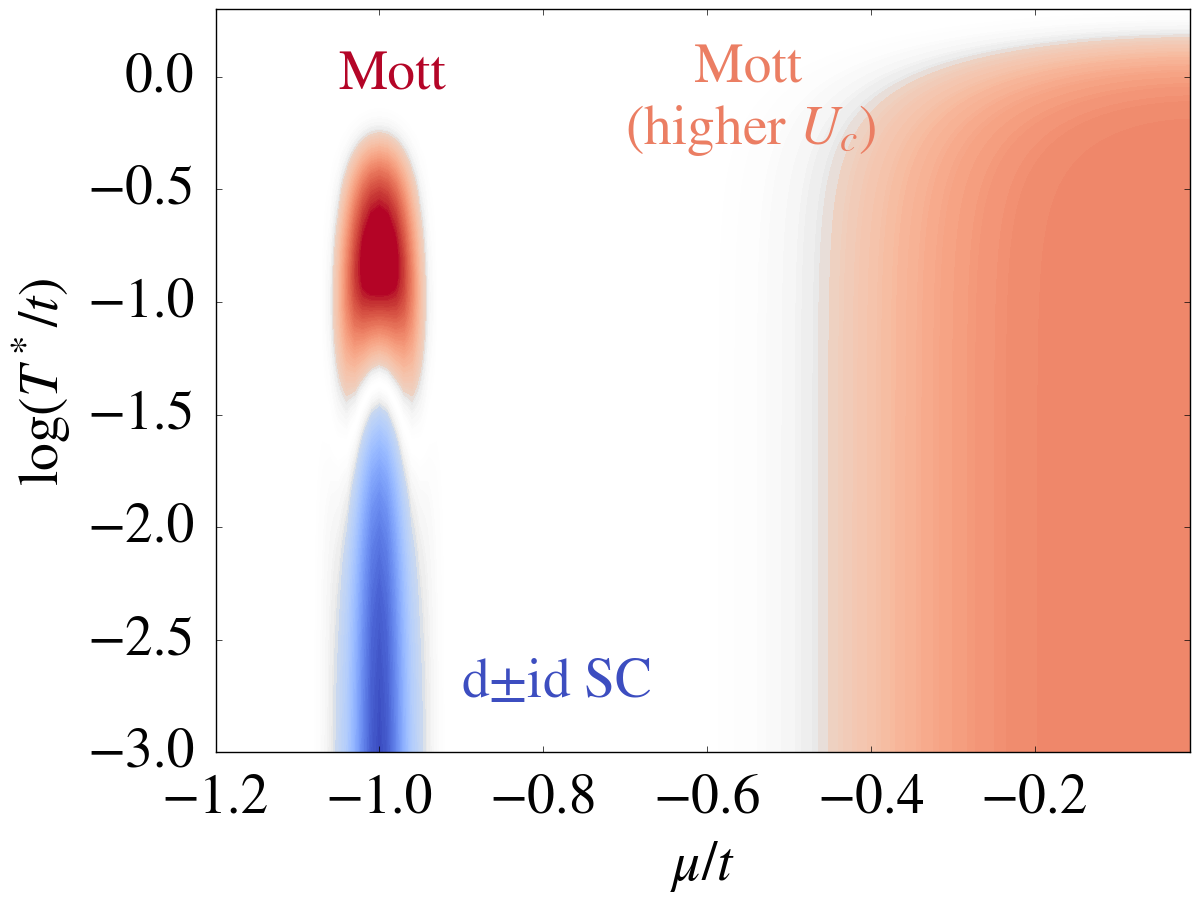}
\caption{Phase diagram of twisted bilayer graphene near the magic angle as a function of the temperature and chemical potential. Interactions $U/t=2$ drive the system into various competing, strongly-correlated phases: Near half-filling of the lower band ($\mu\sim -t$), we observe a crossover between d$\pm$id superconductivity and Mott insulating behaviour as $T$ is increased. Near charge neutrality ($\mu=0$), the system is driven towards a Mott insulator, which, however, features a much higher critical interaction and becomes more pronounced only as $U/t$ is increased (which can be achieved experimentally by tuning the twist angle).}
\label{fig:Phase_diagram}
\end{figure}

To gain insight into the experimental results of Cao et al.~\cite{cao2018correlated, cao2018unconventional}, a wide range of models have been proposed in recent weeks \cite{kang2018}. Without assuming a specific microscoping pairing mechanism, Peltonen and coworkers \cite{peltonen2018mean} used mean-field theory to study SC in TBG and find a strongly inhomogeneous superconducting order parameter. Ray and Das \cite{ray2018wannier} solve the Eliashberg equation for TBG and predict an extended s-wave as the leading pairing symmetry. In contast, Xu and Balents \cite{balents2018}, Zhang \cite{zhang2018low}, Liu et al.~\cite{liu2018}, and (for electron doping) Rademaker and Mellado \cite{rademaker2018} propose a (d+id)-wave pairing symmetry. Using Quantum Monte Carlo, Huang et al.~\cite{huang2018antiferromagnetically} and Guo et al. \cite{guo2018pairing} find a Mott phase for the undoped system and a transition to (d+id) SC at light doping. Similar results are obtained by Fidrysiak et al.~\cite{fidrysiak2018unconventional} using a Gutzwiller approximation. Roy and Juricic \cite{roy2018unconventional} suggest (p+ip) pairing in the superconducting state. Dodaro and coworkers \cite{dodaro2018phases} propose a phase diagram for TBL that contains a nematic phase as well as different superconducting phases. Po and coworkers \cite{po2018} as well as Xu et al.~\cite{lee2018} analyze different insulating states such as intervalley coherent Mott insulator Kekule ordered states, antiferromagnetic insulators, featureless Mott insulators or quantum spin liquids, and outline experiments that can distinguish between these states. Padhi et al.~\cite{padhi2018wigner} suggest that the observed insulating behaviour arises from a Wigner crystal phase. Baskaran \cite{baskaran2018theory} explains SC of TBG in terms of an emergent Josephson-Moir\'e lattice due to resonating valence-bond correlations.

Despite the theoretical progress achieved, quantitative unbiased methods to describe TBG are still sparse. In this paper, we remedy this by studying the effects of strong correlations in TBG near the magic twist angle using a combination of various many-body methods. First, we employ the functional renormalization group (FRG) to determine the effective two-particle interaction $\Gamma_2$ and from that the leading ordering tendency as a function of the temperature and doping. Pictorially, one can think of this approach as a random phase approximation resummation which does not single out one form of two-particle scattering but treats all channels (such as the Cooper or the particle-hole channel) on equal footing. The FRG thus provides an unbiased way to detect competing types of order. The method was successfully applied to study the phase diagram of the $t-t'$ Hubbard model on a square lattice (see \cite{frg2da,frg2db,frg2dc} for early works) as well as of more complex systems \cite{metzner2012functional,wessel2018}. In a second step, we will use a mean-field decoupling to extract pairing symmetries of SC phases.

Our key result is the phase diagram as a function of the temperature and chemical potential $\mu$ shown in Fig.~\ref{fig:Phase_diagram}. At low doping, i.e., near half-filling of the lowest electron bands ($\mu\sim-t$, where $t$ is the prefactor of the kinetic energy), we observe a superconducting dome with a d$\pm$id pairing symmetry which crosses over into a Mott insulating phase as the temperature increases. Near charge neutrality ($\mu=0$), we find a tendency towards forming a Mott insulator; however, the critical interaction strength associated with this phase is higher than at $\mu=-t$. These results are fully consistent with the recent experiments of Cao et al. \cite{cao2018correlated, cao2018unconventional}, where the Mott insulator at $\mu=0$ should show up if $U/t$ is increased by changing the twist angle). The only key difference is that the d$\pm$id superconducting dome around $\mu\sim-t$ is not split by the competing Mott phase occuring at higher temperatures. This could be an artifact of the simplicty of the underlying microscopic model or of our methodology. Our approach, however, can easily be extended to more complex systems, which we defer to an upcoming publication.  

The rest of this paper is devoted to explaining how we obtain the phase diagram shown in Fig.~\ref{fig:Phase_diagram}.

\begin{figure}[t]
\centering
\includegraphics[width=0.95\columnwidth]{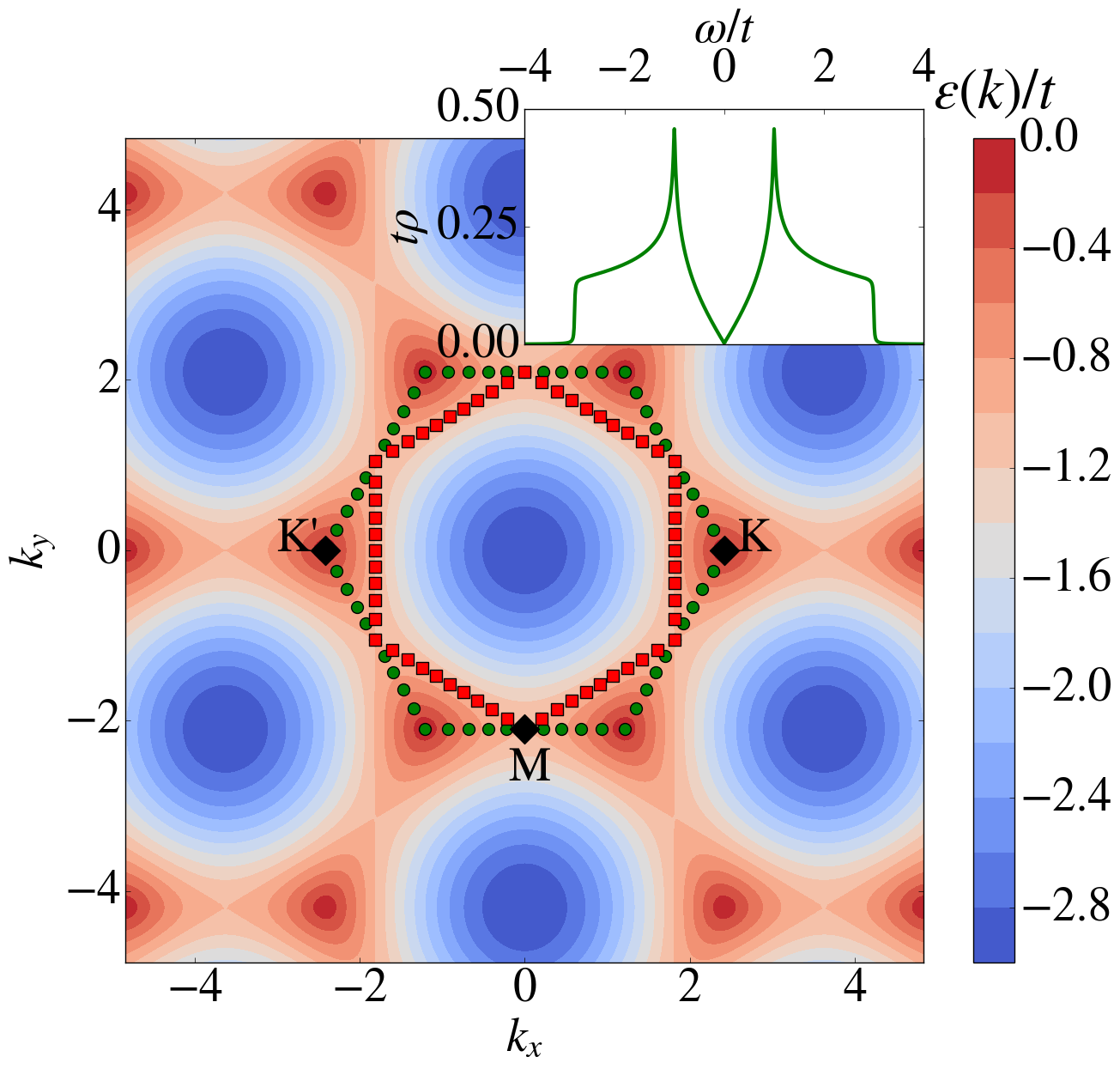}
\caption{Non-interacting dispersion relation of the lower band of twisted bilayer graphene as predicted by the Yuan-Fu model (\ref{eq:h}) \cite{yuan2018model}. The first Brillouin zone is marked by green dots. Near half-filling of the lower band ($\mu\sim-t$), the Fermi surface (red squares) is both highly nested and contains a van Hove singularity in the density of states $\rho$ (shown as the inset).}
\label{fig:Eps}
\end{figure}

\textit{Methods---}
In Ref.~\cite{yuan2018model}, Yuan and Fu used symmetry arguments to construct a tight-binding Hamiltonian which governs the low-energy physics of twisted bilayer graphene in the experimentally-relevant parameter region. The simplest $SU(4)$-symmetric part of their model takes the form of a four-band Hubbard model on a honeycomb lattice,
\begin{equation}\label{eq:h}
H=-t\sum\limits_{\left\langle i,j\right\rangle}\sum\limits_{\sigma=\uparrow,\downarrow\atop p=x,y}\left(c_{i,\sigma,p}^\dagger c_{j,\sigma,p}+{ \rm H.c.}\right)+ U\sum\limits_{i}n_i n_i,
\end{equation} 
where $\sigma$ is the electron spin, and $\{x,y\}$ are two degenerate orbitals (with $p_x$ and $p_y$ symmetry, respectively) located on the triangular sublattices of the honeycomb structure. $c_{i,\sigma,p}$ is the corresponding annihilation operator, and $n_i= \sum_{\sigma=\uparrow,\downarrow}\sum_{p=x,y} c^\dagger_{i,\sigma,p}c_{i,\sigma,p}$. The hopping strength between nearest neighbors and the local Hubbard interaction are denoted by $t$ und $U$, respectively. 

The value of $t$ depends on the twist angle; near the magic angle, $t$ becomes small and thus $U/t$ becomes large. Unless mentioned otherwise, we always work with a fixed $U/t=2$. Note that accounting for next-nearest neighbour hoppings $t_2$ does not qualitatively change our results in general and the phase diagram in particular. We will comment on the inclusion of other terms (e.g., Hund's couplings) below.

We employ a two-step protocol to determine the phase diagram of the Yuan-Fu model. First, we use the functional renormalization group to study the effects of strong correlations induced by $U$. The FRG reformulates this many-body problem in terms of an infinite hierarchy of flow equations for coupling constants with an infrared cutoff $\Lambda$ serving as the flow parameter, see \cite{metzner2012functional} for an introduction. In the context of 2d fermionic systems, one truncates this hierarchy by neglecting the three-particle scattering and focuses solely on the renormalization of the effective two-particle interaction $\Gamma_2^\Lambda(\vec k_1,\vec k_2, \vec k_1',\vec k_2')$ with $\Gamma_2^{\Lambda_\textnormal{initial}}\sim U$. The flow is stopped at a scale $\Lambda_\textnormal{final}$ where this coupling diverges, and the leading ordering tendency can be identified from the momentum structure of $\Gamma_2^{\Lambda_\textnormal{final}}$. In order to solve the flow equation for $\Gamma_2^\Lambda$ in practice, one needs to discretize the Brillouin zone \cite{metzner2012functional}; such technical details about our calculation will be presented elsewhere.

In a second step, we will analyze $\Gamma_2^{\Lambda_\textnormal{final}}$ using a mean-field decoupling. This allows us to extract the pairing symmetry of the superconducting phase.

\textit{Results---}
It is instructive to first discuss the non-interacting band structure of the Yuan-Fu model. At $U=0$, Eq.~(\ref{eq:h}) features particle-hole symmetric upper and lower bands, each with a four-fold (spin and orbital) degeneracy. Figure~\ref{fig:Eps} shows the dispersion relation of the lower (electron) bands; the first Brillouin zone is marked by green dots. The experimentally most interesting regime is near half-filling of the electron bands ($\mu=-t$). In this case, the Fermi surface (red squares) is both highly-nested and contains the van Hove sigularities of the density of states at the M points. At lower values of the doping, there are two Fermi surfaces centered on the K and K' points of the Brillouin
zone, and scattering between these valleys can play an important role. When electron-electron interactions are included, these features of the non-interacting band structure can give rise to different electronic phases such as Mott insulators, superconductivity, or Wigner crystals. This way of understanding the origin of ordering tendencies is well-established for the Hubbard model on a square lattice, for which the FRG succeeds in correctly detecting phases \cite{metzner2012functional}. It is thus resonable to expect that the same holds true for twisted bilayer graphene.

\begin{figure}[t]
\centering
\includegraphics[width=0.95\columnwidth]{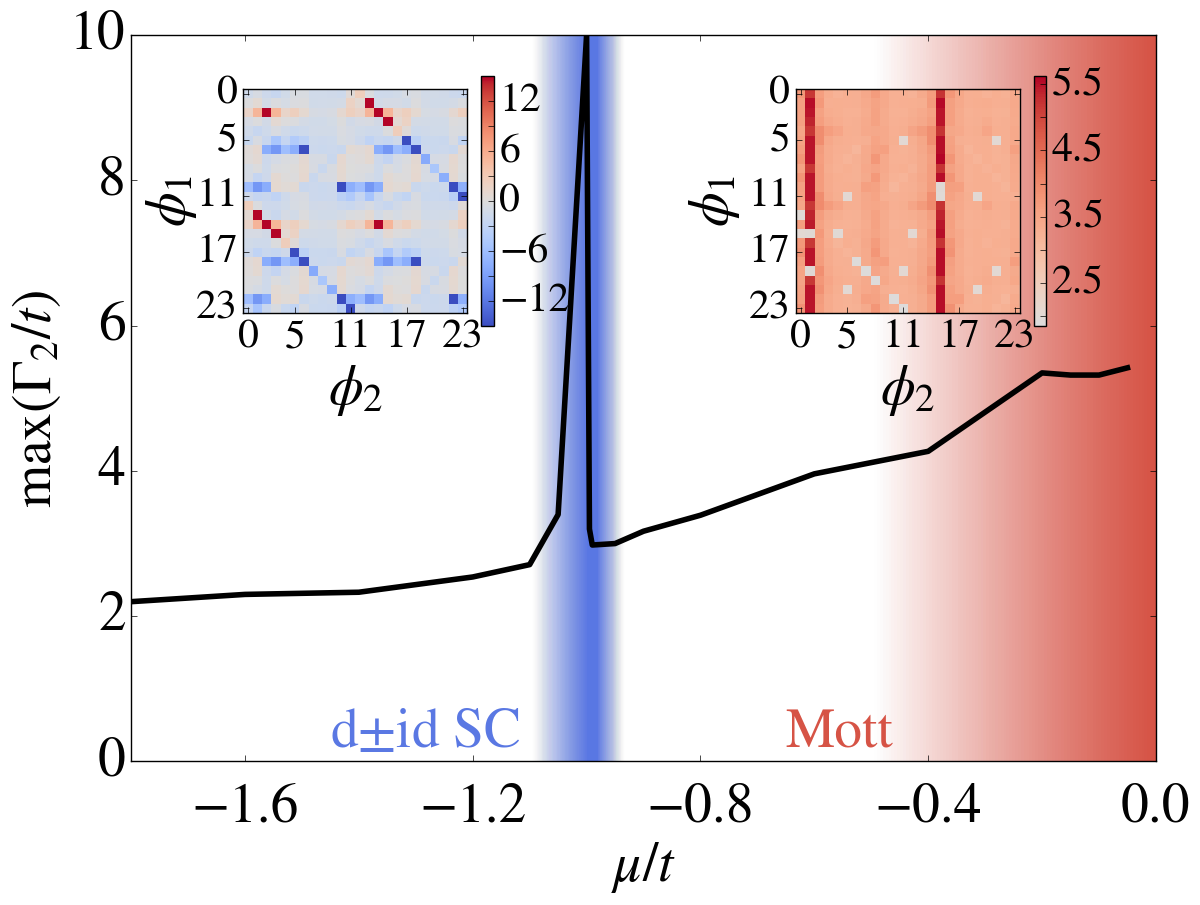}
\caption{Main panel: Maximum value of the effective coupling constant $\Gamma_2^{\Lambda}$  at the end of the RG flow (fixed $\Lambda_\textnormal{final}=10^{-4}t$) as a function of the chemical potential for a bare interaction $U/t=2$. We observe a strong enhancement around $\mu\sim-t$, and the corresponding momentum structure of $\Gamma_2$ along the Fermi surface (left inset) suggests d-wave SC in this region (see the main text for details). Near $\mu=0$, we find a momentum structure that indicates Mott insulating behaviour (right inset). The enhancement of $\Gamma_2$ is much smaller than for the SC but becomes more pronounced when $U/t$ is increased. }
\label{fig:Gam}
\end{figure}

We now use the FRG to study the effects of the electron-electron interactions in the Yuan-Fu model. We first integrate the flow of the two-particle scattering $\Gamma_2^\Lambda$ from $\Lambda_\textnormal{initial}=\infty$ down to a fixed $\Lambda_\textnormal{final}=10^{-4}t$. The maximal absolute value of $\Gamma_2^\Lambda$ at the end of the flow is shown in Fig.~\ref{fig:Gam} as a function of the chemical potential for fixed $U/t=2$. One can see a strong enhancement around half-filling of the electron band ($\mu\sim-t$) as well as a mild enhancement near charge neutrality ($\mu=0$) which becomes more pronounced if $U/t$ is increased (data not shown).

In order to identify the leading ordering tendency around $\mu=0$ and $\mu=-t$, we investigate the momentum-space structure of $\Gamma_2^\Lambda(\vec k_1,\vec k_2, \vec k_1',\vec k_2')$ at $\Lambda_\textnormal{final}$. To this end, we set $\vec k_1'$, $\vec k_1$, and $\vec k_2$ to points on the Fermi surface ($\vec k_2'$ is fixed by momentum conservation), which we paramatrize using a angular variable $\phi$. The insets to Fig.~\ref{fig:Gam} show $\Gamma_2$ for a fixed $\phi_1'$ as a function of the angle of the outgoing momenta $\phi_1$ and $\phi_2$. Near $\mu=-t$ (left inset), we observe diagonal lines $\vec k_1+\vec k_2=0$ with changing signs, indicating a superconducting phase with a d-wave order parameter. This is similar to the physics of the two-dimensional Hubbard model on a square lattice \cite{metzner2012functional}. The dominant pairing occurs between  $(\uparrow,x,\text{lower band})$ and $(\downarrow,y,\text{lower band})$ (and all symmetry-related pairs), which is a superconducting pairing between particles with opposing quantum numbers in the electron band. In the vicinity of $\mu=0$ (right inset), the momentum structure of vertex looks profoundly different: It features vertical lines, which is again reminiscent of the Mott insulating state in the two-dimensional Hubbard model on a square lattice \cite{metzner2012functional}. The dominant pairing in this regime occurs between $(\uparrow,x,\text{upper band})$ and  $(\downarrow,y,\text{lower band})$, which minimizes the kinetic energy.

In order to establish the phase diagram shown in Fig.~\ref{fig:Phase_diagram}, we monitor the FRG flow as a function of $\Lambda$, which we define as an effective temperature $T^*$. If the maximal value of $\Gamma_2^\Lambda$ stays below a pre-defined threshold $U_c$, we interpret this as a metallic phase; if it exceeds $U_c$, we determine the corresponding type of order by looking at the momentum structure of $\Gamma_2^{\Lambda}$. Since for $U/t=2$ the enhancement of $\Gamma_2$ around $\mu=0$ is only mild (see Fig.~\ref{fig:Gam}), we choose a rather small $U_c/t=2.8$ (alternatively, one could work with a larger bare $U/t$). With this definition, we obtain the phase diagram shown in Fig.~\ref{fig:Phase_diagram}. It is important to stress that -- while our results are based on a quantitative many-body calculation -- Fig.~\ref{fig:Phase_diagram} is only correct on a qualitative level due to the arbitrariness of our choice of $U_c$, the simplicity of our model, and the approximations underlying our approach \cite{metzner2012functional}.

\begin{figure}[t]
\centering
\includegraphics[width=0.9\columnwidth]{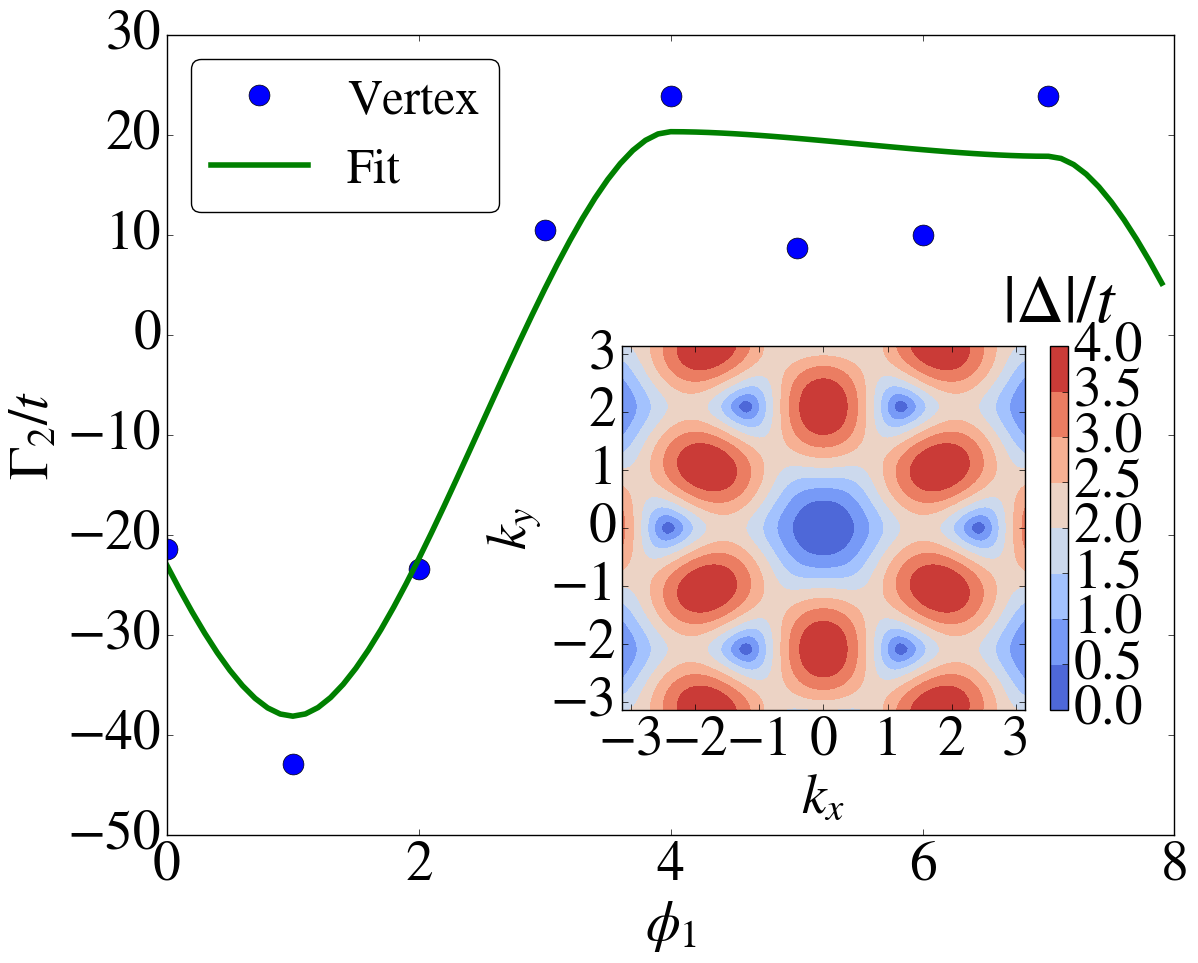}
\caption{Main panel: Fit of the effective coupling constant $\Gamma_2$ near $\mu=-t$ along the dominant diagonal $\vec k_1+\vec k_2=0$ to the form factors $d_{x^2-y^2}$ and $d_{xy}$; the prefactors are found to be $\sim 4.5$  and $\sim -14.6$, respectively. Inset: Absolute value of the gap function (modulo an overall factor) for the mean-field solution that minimizes the minimal grand canonical potential. The gap is large at the van Hove points (which minimizes the grand free energy). }
\label{fig:fitandgap}
\end{figure}

We finally analyze the superconducting phase that occurs near $\mu=-t$ in more detail in order to determine the precise form of the pairing symmetry. We parametrize the effective interaction at $\Lambda_\textnormal{final}$ in a way that reflects the pronounced diagonal structure along $\vec k_1+\vec k_2=0$,
\begin{equation}
\Gamma_2^\Lambda(\vec k_1,\vec k_2,\vec k_1',\vec k_2')=\Gamma_2(\vec k_1,\vec k_1') \delta_{\vec k_1,-\vec k_2}\delta_{\vec k_1',-\vec k_2'},
\end{equation}
where $\Gamma_2(\vec k,\vec k')=a d_{x^2-y^2}(\vec k)d_{x^2-y^2}(\vec k')+bd_{xy}(\vec k)d_{xy}(\vec k')$, and $d_{xy}$ as well as $d_{x^2-y^2}$ denote the form factors of the superconducting order parameter \cite{kiesel2012competing}.  The coefficients $a$ and $b$ are then determined by fitting, see Fig.~\ref{fig:fitandgap}. If we now insert this vertex into the BCS mean-field equation for the superconducting order parameter $\Delta_{\vec q}$, one can immediately see that $\Delta_{\vec q}$ must have the same functional form:
\begin{align}
\Delta_{\vec q}&=-\frac{1}{N}\sum\limits_{\vec k} \Gamma_2(\vec k,\vec q) \frac{\Delta_{\vec k}}{2E(\vec k)}\tanh\left(\frac{E(\vec k)}{2T}\right)\notag\\
&=c_1 d_{x^2-y^2}(\vec q)+ c_2e^{i\phi}d_{xy}(\vec q).
\end{align}
The phase $\phi$ is not determined by our FRG calculation but can be extracted by minimizing the grand-canonical potential $\Omega$. Instead of resorting to numerics, we employ a simple argument valid at $\mu=-t$. If we assume that $\Omega$ is dominated by momenta on the Fermi surface in general and by the van Hove singularities $\vec k_{\rm vH}$ in particular (which lie on the Fermi surface for $\mu=-t$), we obtain \cite{platt2012}
\begin{equation}
\Omega\sim-\sum\limits_{\vec q=\vec k_{\rm vH}} \left|\Delta_{\vec q}(\phi)\right|.
\end{equation}
This expression is minimized by $\phi=\pm \pi/2$ for arbitrary $c_{1,2}$, and the superconducting phase near half-filling of the electron band thus features a d$\pm$id pairing symmetry. The corresponding gap function $\Delta_{\vec q}$ is shown in the inset to Fig.~\ref{fig:fitandgap}; its absolute value is maximal at the van Hove points.

\textit{Conclusions---}
We have reported the phase diagram of twisted bilayer graphene near the magic twist angle by studying the effects of strong correlations within the effective low-energy model devised by Yuan and Fu \cite{yuan2018model}. We used the functional renomalization group -- a method which can reliably detect ordering tendencies of interacting 2d systems such as the Hubbard model on a square lattice \cite{metzner2012functional} -- combined with a mean-field analysis of the effective two-particle interactions at the end of the FRG flow. Near half-filling of the electron band, we found d$\pm$id superconductivity crossing over to a Mott insulator as the temperature increases. Near charge neutrality, we detected a weaker tendency to form a Mott insulator. Our results provide an unbiased frst step towards explaining recent experiments on twisted bilayer graphene \cite{cao2018correlated,cao2018unconventional} and establish correlations as the origin of the phenomena they observe.

As a next step, one should study generalizations of the Yuan-Fu model by adding, e.g., Hund's couplings. It would also be interesting to directly work with ab initio band structures \cite{trambly2010localization} and to investigate the effects of long-ranged screened Coulomb interactions \cite{stauber2016quasi}. All of this is straighforward within our approach but left for future work. 

\textit{Acknowledgements---}
We thank Fabiano Corsetti, Jannis Ehrlich, Julian Lichtenstein, and Arash Mostofi for very fruitful discussions and in particular Carsten Honerkamp for sharing technical details of a 2d FRG implementation. DMK and CK acknowledge support by the Deutsche Forschungsgemeinschaft through the Emmy Noether program (KA 3360/2-1).

\bibliography{FRGTBG}

\begin{thebibliography}{32}%
\makeatletter
\providecommand \@ifxundefined [1]{%
 \@ifx{#1\undefined}
}%
\providecommand \@ifnum [1]{%
 \ifnum #1\expandafter \@firstoftwo
 \else \expandafter \@secondoftwo
 \fi
}%
\providecommand \@ifx [1]{%
 \ifx #1\expandafter \@firstoftwo
 \else \expandafter \@secondoftwo
 \fi
}%
\providecommand \natexlab [1]{#1}%
\providecommand \enquote  [1]{``#1''}%
\providecommand \bibnamefont  [1]{#1}%
\providecommand \bibfnamefont [1]{#1}%
\providecommand \citenamefont [1]{#1}%
\providecommand \href@noop [0]{\@secondoftwo}%
\providecommand \href [0]{\begingroup \@sanitize@url \@href}%
\providecommand \@href[1]{\@@startlink{#1}\@@href}%
\providecommand \@@href[1]{\endgroup#1\@@endlink}%
\providecommand \@sanitize@url [0]{\catcode `\\12\catcode `\$12\catcode
  `\&12\catcode `\#12\catcode `\^12\catcode `\_12\catcode `\%12\relax}%
\providecommand \@@startlink[1]{}%
\providecommand \@@endlink[0]{}%
\providecommand \url  [0]{\begingroup\@sanitize@url \@url }%
\providecommand \@url [1]{\endgroup\@href {#1}{\urlprefix }}%
\providecommand \urlprefix  [0]{URL }%
\providecommand \Eprint [0]{\href }%
\providecommand \doibase [0]{http://dx.doi.org/}%
\providecommand \selectlanguage [0]{\@gobble}%
\providecommand \bibinfo  [0]{\@secondoftwo}%
\providecommand \bibfield  [0]{\@secondoftwo}%
\providecommand \translation [1]{[#1]}%
\providecommand \BibitemOpen [0]{}%
\providecommand \bibitemStop [0]{}%
\providecommand \bibitemNoStop [0]{.\EOS\space}%
\providecommand \EOS [0]{\spacefactor3000\relax}%
\providecommand \BibitemShut  [1]{\csname bibitem#1\endcsname}%
\let\auto@bib@innerbib\@empty
\bibitem [{\citenamefont {Cao}\ \emph {et~al.}(2018{\natexlab{a}})\citenamefont
  {Cao}, \citenamefont {Fatemi}, \citenamefont {Demir}, \citenamefont {Fang},
  \citenamefont {Tomarken}, \citenamefont {Luo}, \citenamefont
  {Sanchez-Yamagishi}, \citenamefont {Watanabe}, \citenamefont {Taniguchi},
  \citenamefont {Kaxiras} \emph {et~al.}}]{cao2018correlated}%
  \BibitemOpen
  \bibfield  {author} {\bibinfo {author} {\bibfnamefont {Y.}~\bibnamefont
  {Cao}}, \bibinfo {author} {\bibfnamefont {V.}~\bibnamefont {Fatemi}},
  \bibinfo {author} {\bibfnamefont {A.}~\bibnamefont {Demir}}, \bibinfo
  {author} {\bibfnamefont {S.}~\bibnamefont {Fang}}, \bibinfo {author}
  {\bibfnamefont {S.~L.}\ \bibnamefont {Tomarken}}, \bibinfo {author}
  {\bibfnamefont {J.~Y.}\ \bibnamefont {Luo}}, \bibinfo {author} {\bibfnamefont
  {J.~D.}\ \bibnamefont {Sanchez-Yamagishi}}, \bibinfo {author} {\bibfnamefont
  {K.}~\bibnamefont {Watanabe}}, \bibinfo {author} {\bibfnamefont
  {T.}~\bibnamefont {Taniguchi}}, \bibinfo {author} {\bibfnamefont
  {E.}~\bibnamefont {Kaxiras}},  \emph {et~al.},\ }\href@noop {} {\bibfield
  {journal} {\bibinfo  {journal} {Nature}\ }\textbf {\bibinfo {volume} {556}},\
  \bibinfo {pages} {80} (\bibinfo {year} {2018}{\natexlab{a}})}\BibitemShut
  {NoStop}%
\bibitem [{\citenamefont {Cao}\ \emph {et~al.}(2018{\natexlab{b}})\citenamefont
  {Cao}, \citenamefont {Fatemi}, \citenamefont {Fang}, \citenamefont
  {Watanabe}, \citenamefont {Taniguchi}, \citenamefont {Kaxiras},\ and\
  \citenamefont {Jarillo-Herrero}}]{cao2018unconventional}%
  \BibitemOpen
  \bibfield  {author} {\bibinfo {author} {\bibfnamefont {Y.}~\bibnamefont
  {Cao}}, \bibinfo {author} {\bibfnamefont {V.}~\bibnamefont {Fatemi}},
  \bibinfo {author} {\bibfnamefont {S.}~\bibnamefont {Fang}}, \bibinfo {author}
  {\bibfnamefont {K.}~\bibnamefont {Watanabe}}, \bibinfo {author}
  {\bibfnamefont {T.}~\bibnamefont {Taniguchi}}, \bibinfo {author}
  {\bibfnamefont {E.}~\bibnamefont {Kaxiras}}, \ and\ \bibinfo {author}
  {\bibfnamefont {P.}~\bibnamefont {Jarillo-Herrero}},\ }\href@noop {}
  {\bibfield  {journal} {\bibinfo  {journal} {Nature}\ }\textbf {\bibinfo
  {volume} {556}},\ \bibinfo {pages} {43} (\bibinfo {year}
  {2018}{\natexlab{b}})}\BibitemShut {NoStop}%
\bibitem [{\citenamefont {Chung}\ \emph {et~al.}(2018)\citenamefont {Chung},
  \citenamefont {Xu},\ and\ \citenamefont {Chen}}]{chung2018}%
  \BibitemOpen
  \bibfield  {author} {\bibinfo {author} {\bibfnamefont {T.-F.}\ \bibnamefont
  {Chung}}, \bibinfo {author} {\bibfnamefont {Y.}~\bibnamefont {Xu}}, \ and\
  \bibinfo {author} {\bibfnamefont {Y.~P.}\ \bibnamefont {Chen}},\ }\href@noop
  {} {\bibfield  {journal} {\bibinfo  {journal} {arXiv preprint
  arXiv:1805.01454}\ } (\bibinfo {year} {2018})}\BibitemShut {NoStop}%
\bibitem [{\citenamefont {Qiao}\ and\ \citenamefont {He}(2018)}]{qiao2018}%
  \BibitemOpen
  \bibfield  {author} {\bibinfo {author} {\bibfnamefont {J.-B.}\ \bibnamefont
  {Qiao}}\ and\ \bibinfo {author} {\bibfnamefont {L.}~\bibnamefont {He}},\
  }\href@noop {} {\bibfield  {journal} {\bibinfo  {journal} {arXiv preprint
  arXiv:1805.03790}\ } (\bibinfo {year} {2018})}\BibitemShut {NoStop}%
\bibitem [{\citenamefont {Bistritzer}\ and\ \citenamefont
  {MacDonald}(2011)}]{bistritzer2011moire}%
  \BibitemOpen
  \bibfield  {author} {\bibinfo {author} {\bibfnamefont {R.}~\bibnamefont
  {Bistritzer}}\ and\ \bibinfo {author} {\bibfnamefont {A.~H.}\ \bibnamefont
  {MacDonald}},\ }\href@noop {} {\bibfield  {journal} {\bibinfo  {journal}
  {Proceedings of the National Academy of Sciences}\ }\textbf {\bibinfo
  {volume} {108}},\ \bibinfo {pages} {12233} (\bibinfo {year}
  {2011})}\BibitemShut {NoStop}%
\bibitem [{\citenamefont {dos Santos}\ \emph {et~al.}(2012)\citenamefont {dos
  Santos}, \citenamefont {Peres},\ and\ \citenamefont
  {Neto}}]{dos2012continuum}%
  \BibitemOpen
  \bibfield  {author} {\bibinfo {author} {\bibfnamefont {J.~L.}\ \bibnamefont
  {dos Santos}}, \bibinfo {author} {\bibfnamefont {N.}~\bibnamefont {Peres}}, \
  and\ \bibinfo {author} {\bibfnamefont {A.~C.}\ \bibnamefont {Neto}},\
  }\href@noop {} {\bibfield  {journal} {\bibinfo  {journal} {Physical Review
  B}\ }\textbf {\bibinfo {volume} {86}},\ \bibinfo {pages} {155449} (\bibinfo
  {year} {2012})}\BibitemShut {NoStop}%
\bibitem [{\citenamefont {Kang}\ and\ \citenamefont {Vafek}(2018)}]{kang2018}%
  \BibitemOpen
  \bibfield  {author} {\bibinfo {author} {\bibfnamefont {J.}~\bibnamefont
  {Kang}}\ and\ \bibinfo {author} {\bibfnamefont {O.}~\bibnamefont {Vafek}},\
  }\href@noop {} {\bibfield  {journal} {\bibinfo  {journal} {arXiv preprint
  arXiv:1805.04918}\ } (\bibinfo {year} {2018})}\BibitemShut {NoStop}%
\bibitem [{\citenamefont {Peltonen}\ \emph {et~al.}(2018)\citenamefont
  {Peltonen}, \citenamefont {Ojaj{\"a}rvi},\ and\ \citenamefont
  {Heikkil{\"a}}}]{peltonen2018mean}%
  \BibitemOpen
  \bibfield  {author} {\bibinfo {author} {\bibfnamefont {T.~J.}\ \bibnamefont
  {Peltonen}}, \bibinfo {author} {\bibfnamefont {R.}~\bibnamefont
  {Ojaj{\"a}rvi}}, \ and\ \bibinfo {author} {\bibfnamefont {T.~T.}\
  \bibnamefont {Heikkil{\"a}}},\ }\href@noop {} {\bibfield  {journal} {\bibinfo
   {journal} {arXiv preprint arXiv:1805.01039}\ } (\bibinfo {year}
  {2018})}\BibitemShut {NoStop}%
\bibitem [{\citenamefont {Ray}\ and\ \citenamefont
  {Das}(2018)}]{ray2018wannier}%
  \BibitemOpen
  \bibfield  {author} {\bibinfo {author} {\bibfnamefont {S.}~\bibnamefont
  {Ray}}\ and\ \bibinfo {author} {\bibfnamefont {T.}~\bibnamefont {Das}},\
  }\href@noop {} {\bibfield  {journal} {\bibinfo  {journal} {arXiv preprint
  arXiv:1804.09674}\ } (\bibinfo {year} {2018})}\BibitemShut {NoStop}%
\bibitem [{\citenamefont {Xu}\ and\ \citenamefont
  {Balents}(2018)}]{balents2018}%
  \BibitemOpen
  \bibfield  {author} {\bibinfo {author} {\bibfnamefont {C.}~\bibnamefont
  {Xu}}\ and\ \bibinfo {author} {\bibfnamefont {L.}~\bibnamefont {Balents}},\
  }\href@noop {} {\bibfield  {journal} {\bibinfo  {journal} {arXiv preprint
  arXiv:1803.08057}\ } (\bibinfo {year} {2018})}\BibitemShut {NoStop}%
\bibitem [{\citenamefont {Zhang}(2018)}]{zhang2018low}%
  \BibitemOpen
  \bibfield  {author} {\bibinfo {author} {\bibfnamefont {L.}~\bibnamefont
  {Zhang}},\ }\href@noop {} {\bibfield  {journal} {\bibinfo  {journal} {arXiv
  preprint arXiv:1804.09047}\ } (\bibinfo {year} {2018})}\BibitemShut {NoStop}%
\bibitem [{\citenamefont {Liu}\ \emph {et~al.}(2018)\citenamefont {Liu},
  \citenamefont {Zhang}, \citenamefont {Chen},\ and\ \citenamefont
  {Yang}}]{liu2018}%
  \BibitemOpen
  \bibfield  {author} {\bibinfo {author} {\bibfnamefont {C.-C.}\ \bibnamefont
  {Liu}}, \bibinfo {author} {\bibfnamefont {L.-D.}\ \bibnamefont {Zhang}},
  \bibinfo {author} {\bibfnamefont {W.-Q.}\ \bibnamefont {Chen}}, \ and\
  \bibinfo {author} {\bibfnamefont {F.}~\bibnamefont {Yang}},\ }\href@noop {}
  {\bibfield  {journal} {\bibinfo  {journal} {arXiv preprint arXiv:1804.10009}\
  } (\bibinfo {year} {2018})}\BibitemShut {NoStop}%
\bibitem [{\citenamefont {Rademaker}\ and\ \citenamefont
  {Mellado}(2018)}]{rademaker2018}%
  \BibitemOpen
  \bibfield  {author} {\bibinfo {author} {\bibfnamefont {L.}~\bibnamefont
  {Rademaker}}\ and\ \bibinfo {author} {\bibfnamefont {P.}~\bibnamefont
  {Mellado}},\ }\href@noop {} {\bibfield  {journal} {\bibinfo  {journal} {arXiv
  preprint arXiv:1805.05294}\ } (\bibinfo {year} {2018})}\BibitemShut {NoStop}%
\bibitem [{\citenamefont {Huang}\ \emph {et~al.}(2018)\citenamefont {Huang},
  \citenamefont {Zhang},\ and\ \citenamefont
  {Ma}}]{huang2018antiferromagnetically}%
  \BibitemOpen
  \bibfield  {author} {\bibinfo {author} {\bibfnamefont {T.}~\bibnamefont
  {Huang}}, \bibinfo {author} {\bibfnamefont {L.}~\bibnamefont {Zhang}}, \ and\
  \bibinfo {author} {\bibfnamefont {T.}~\bibnamefont {Ma}},\ }\href@noop {}
  {\bibfield  {journal} {\bibinfo  {journal} {arXiv preprint arXiv:1804.06096}\
  } (\bibinfo {year} {2018})}\BibitemShut {NoStop}%
\bibitem [{\citenamefont {Guo}\ \emph {et~al.}(2018)\citenamefont {Guo},
  \citenamefont {Zhu}, \citenamefont {Feng},\ and\ \citenamefont
  {Scalettar}}]{guo2018pairing}%
  \BibitemOpen
  \bibfield  {author} {\bibinfo {author} {\bibfnamefont {H.}~\bibnamefont
  {Guo}}, \bibinfo {author} {\bibfnamefont {X.}~\bibnamefont {Zhu}}, \bibinfo
  {author} {\bibfnamefont {S.}~\bibnamefont {Feng}}, \ and\ \bibinfo {author}
  {\bibfnamefont {R.~T.}\ \bibnamefont {Scalettar}},\ }\href@noop {} {\bibfield
   {journal} {\bibinfo  {journal} {arXiv preprint arXiv:1804.00159}\ }
  (\bibinfo {year} {2018})}\BibitemShut {NoStop}%
\bibitem [{\citenamefont {Fidrysiak}\ \emph {et~al.}(2018)\citenamefont
  {Fidrysiak}, \citenamefont {Zegrodnik},\ and\ \citenamefont
  {Spa{\l}ek}}]{fidrysiak2018unconventional}%
  \BibitemOpen
  \bibfield  {author} {\bibinfo {author} {\bibfnamefont {M.}~\bibnamefont
  {Fidrysiak}}, \bibinfo {author} {\bibfnamefont {M.}~\bibnamefont
  {Zegrodnik}}, \ and\ \bibinfo {author} {\bibfnamefont {J.}~\bibnamefont
  {Spa{\l}ek}},\ }\href@noop {} {\bibfield  {journal} {\bibinfo  {journal}
  {arXiv preprint arXiv:1805.01179}\ } (\bibinfo {year} {2018})}\BibitemShut
  {NoStop}%
\bibitem [{\citenamefont {Roy}\ and\ \citenamefont
  {Juricic}(2018)}]{roy2018unconventional}%
  \BibitemOpen
  \bibfield  {author} {\bibinfo {author} {\bibfnamefont {B.}~\bibnamefont
  {Roy}}\ and\ \bibinfo {author} {\bibfnamefont {V.}~\bibnamefont {Juricic}},\
  }\href@noop {} {\bibfield  {journal} {\bibinfo  {journal} {arXiv preprint
  arXiv:1803.11190}\ } (\bibinfo {year} {2018})}\BibitemShut {NoStop}%
\bibitem [{\citenamefont {Dodaro}\ \emph {et~al.}(2018)\citenamefont {Dodaro},
  \citenamefont {Kivelson}, \citenamefont {Schattner}, \citenamefont {Sun},\
  and\ \citenamefont {Wang}}]{dodaro2018phases}%
  \BibitemOpen
  \bibfield  {author} {\bibinfo {author} {\bibfnamefont {J.~F.}\ \bibnamefont
  {Dodaro}}, \bibinfo {author} {\bibfnamefont {S.~A.}\ \bibnamefont
  {Kivelson}}, \bibinfo {author} {\bibfnamefont {Y.}~\bibnamefont {Schattner}},
  \bibinfo {author} {\bibfnamefont {X.-Q.}\ \bibnamefont {Sun}}, \ and\
  \bibinfo {author} {\bibfnamefont {C.}~\bibnamefont {Wang}},\ }\href@noop {}
  {\bibfield  {journal} {\bibinfo  {journal} {arXiv preprint arXiv:1804.03162}\
  } (\bibinfo {year} {2018})}\BibitemShut {NoStop}%
\bibitem [{\citenamefont {Po}\ \emph {et~al.}(2018)\citenamefont {Po},
  \citenamefont {Zou}, \citenamefont {Vishwanath},\ and\ \citenamefont
  {Senthil}}]{po2018}%
  \BibitemOpen
  \bibfield  {author} {\bibinfo {author} {\bibfnamefont {H.~C.}\ \bibnamefont
  {Po}}, \bibinfo {author} {\bibfnamefont {L.}~\bibnamefont {Zou}}, \bibinfo
  {author} {\bibfnamefont {A.}~\bibnamefont {Vishwanath}}, \ and\ \bibinfo
  {author} {\bibfnamefont {T.}~\bibnamefont {Senthil}},\ }\href@noop {}
  {\bibfield  {journal} {\bibinfo  {journal} {arXiv preprint arXiv:1803.09742}\
  } (\bibinfo {year} {2018})}\BibitemShut {NoStop}%
\bibitem [{\citenamefont {Xu}\ \emph {et~al.}(2018)\citenamefont {Xu},
  \citenamefont {Law},\ and\ \citenamefont {Lee}}]{lee2018}%
  \BibitemOpen
  \bibfield  {author} {\bibinfo {author} {\bibfnamefont {X.~Y.}\ \bibnamefont
  {Xu}}, \bibinfo {author} {\bibfnamefont {K.~T.}\ \bibnamefont {Law}}, \ and\
  \bibinfo {author} {\bibfnamefont {P.~A.}\ \bibnamefont {Lee}},\ }\href@noop
  {} {\bibfield  {journal} {\bibinfo  {journal} {arXiv preprint
  arXiv:1805.00478}\ } (\bibinfo {year} {2018})}\BibitemShut {NoStop}%
\bibitem [{\citenamefont {Padhi}\ \emph {et~al.}(2018)\citenamefont {Padhi},
  \citenamefont {Setty},\ and\ \citenamefont {Phillips}}]{padhi2018wigner}%
  \BibitemOpen
  \bibfield  {author} {\bibinfo {author} {\bibfnamefont {B.}~\bibnamefont
  {Padhi}}, \bibinfo {author} {\bibfnamefont {C.}~\bibnamefont {Setty}}, \ and\
  \bibinfo {author} {\bibfnamefont {P.~W.}\ \bibnamefont {Phillips}},\
  }\href@noop {} {\bibfield  {journal} {\bibinfo  {journal} {arXiv preprint
  arXiv:1804.01101}\ } (\bibinfo {year} {2018})}\BibitemShut {NoStop}%
\bibitem [{\citenamefont {Baskaran}(2018)}]{baskaran2018theory}%
  \BibitemOpen
  \bibfield  {author} {\bibinfo {author} {\bibfnamefont {G.}~\bibnamefont
  {Baskaran}},\ }\href@noop {} {\bibfield  {journal} {\bibinfo  {journal}
  {arXiv preprint arXiv:1804.00627}\ } (\bibinfo {year} {2018})}\BibitemShut
  {NoStop}%
\bibitem [{\citenamefont {Zanchi}\ and\ \citenamefont {Schulz}(1998)}]{frg2da}%
  \BibitemOpen
  \bibfield  {author} {\bibinfo {author} {\bibfnamefont {D.}~\bibnamefont
  {Zanchi}}\ and\ \bibinfo {author} {\bibfnamefont {H.~J.}\ \bibnamefont
  {Schulz}},\ }\href {http://stacks.iop.org/0295-5075/44/i=2/a=235} {\bibfield
  {journal} {\bibinfo  {journal} {EPL (Europhysics Letters)}\ }\textbf
  {\bibinfo {volume} {44}},\ \bibinfo {pages} {235} (\bibinfo {year}
  {1998})}\BibitemShut {NoStop}%
\bibitem [{\citenamefont {Halboth}\ and\ \citenamefont
  {Metzner}(2000)}]{frg2db}%
  \BibitemOpen
  \bibfield  {author} {\bibinfo {author} {\bibfnamefont {C.~J.}\ \bibnamefont
  {Halboth}}\ and\ \bibinfo {author} {\bibfnamefont {W.}~\bibnamefont
  {Metzner}},\ }\href {\doibase 10.1103/PhysRevB.61.7364} {\bibfield  {journal}
  {\bibinfo  {journal} {Phys. Rev. B}\ }\textbf {\bibinfo {volume} {61}},\
  \bibinfo {pages} {7364} (\bibinfo {year} {2000})}\BibitemShut {NoStop}%
\bibitem [{\citenamefont {Honerkamp}\ \emph {et~al.}(2001)\citenamefont
  {Honerkamp}, \citenamefont {Salmhofer}, \citenamefont {Furukawa},\ and\
  \citenamefont {Rice}}]{frg2dc}%
  \BibitemOpen
  \bibfield  {author} {\bibinfo {author} {\bibfnamefont {C.}~\bibnamefont
  {Honerkamp}}, \bibinfo {author} {\bibfnamefont {M.}~\bibnamefont
  {Salmhofer}}, \bibinfo {author} {\bibfnamefont {N.}~\bibnamefont {Furukawa}},
  \ and\ \bibinfo {author} {\bibfnamefont {T.~M.}\ \bibnamefont {Rice}},\
  }\href {\doibase 10.1103/PhysRevB.63.035109} {\bibfield  {journal} {\bibinfo
  {journal} {Phys. Rev. B}\ }\textbf {\bibinfo {volume} {63}},\ \bibinfo
  {pages} {035109} (\bibinfo {year} {2001})}\BibitemShut {NoStop}%
\bibitem [{\citenamefont {Metzner}\ \emph {et~al.}(2012)\citenamefont
  {Metzner}, \citenamefont {Salmhofer}, \citenamefont {Honerkamp},
  \citenamefont {Meden},\ and\ \citenamefont
  {Sch{\"o}nhammer}}]{metzner2012functional}%
  \BibitemOpen
  \bibfield  {author} {\bibinfo {author} {\bibfnamefont {W.}~\bibnamefont
  {Metzner}}, \bibinfo {author} {\bibfnamefont {M.}~\bibnamefont {Salmhofer}},
  \bibinfo {author} {\bibfnamefont {C.}~\bibnamefont {Honerkamp}}, \bibinfo
  {author} {\bibfnamefont {V.}~\bibnamefont {Meden}}, \ and\ \bibinfo {author}
  {\bibfnamefont {K.}~\bibnamefont {Sch{\"o}nhammer}},\ }\href@noop {}
  {\bibfield  {journal} {\bibinfo  {journal} {Reviews of Modern Physics}\
  }\textbf {\bibinfo {volume} {84}},\ \bibinfo {pages} {299} (\bibinfo {year}
  {2012})}\BibitemShut {NoStop}%
\bibitem [{\citenamefont {Hesselmann}\ \emph {et~al.}(2018)\citenamefont
  {Hesselmann}, \citenamefont {Scherer}, \citenamefont {Scherer},\ and\
  \citenamefont {Wessel}}]{wessel2018}%
  \BibitemOpen
  \bibfield  {author} {\bibinfo {author} {\bibfnamefont {S.}~\bibnamefont
  {Hesselmann}}, \bibinfo {author} {\bibfnamefont {D.~D.}\ \bibnamefont
  {Scherer}}, \bibinfo {author} {\bibfnamefont {M.~M.}\ \bibnamefont
  {Scherer}}, \ and\ \bibinfo {author} {\bibfnamefont {S.}~\bibnamefont
  {Wessel}},\ }\href@noop {} {\bibfield  {journal} {\bibinfo  {journal} {arXiv
  preprint arXiv:1804.11131}\ } (\bibinfo {year} {2018})}\BibitemShut {NoStop}%
\bibitem [{\citenamefont {Yuan}\ and\ \citenamefont
  {Fu}(2018)}]{yuan2018model}%
  \BibitemOpen
  \bibfield  {author} {\bibinfo {author} {\bibfnamefont {N.~F.}\ \bibnamefont
  {Yuan}}\ and\ \bibinfo {author} {\bibfnamefont {L.}~\bibnamefont {Fu}},\
  }\href@noop {} {\bibfield  {journal} {\bibinfo  {journal} {arXiv preprint
  arXiv:1803.09699}\ } (\bibinfo {year} {2018})}\BibitemShut {NoStop}%
\bibitem [{\citenamefont {Kiesel}\ \emph {et~al.}(2012)\citenamefont {Kiesel},
  \citenamefont {Platt}, \citenamefont {Hanke}, \citenamefont {Abanin},\ and\
  \citenamefont {Thomale}}]{kiesel2012competing}%
  \BibitemOpen
  \bibfield  {author} {\bibinfo {author} {\bibfnamefont {M.~L.}\ \bibnamefont
  {Kiesel}}, \bibinfo {author} {\bibfnamefont {C.}~\bibnamefont {Platt}},
  \bibinfo {author} {\bibfnamefont {W.}~\bibnamefont {Hanke}}, \bibinfo
  {author} {\bibfnamefont {D.~A.}\ \bibnamefont {Abanin}}, \ and\ \bibinfo
  {author} {\bibfnamefont {R.}~\bibnamefont {Thomale}},\ }\href@noop {}
  {\bibfield  {journal} {\bibinfo  {journal} {Physical Review B}\ }\textbf
  {\bibinfo {volume} {86}},\ \bibinfo {pages} {020507} (\bibinfo {year}
  {2012})}\BibitemShut {NoStop}%
\bibitem [{\citenamefont {Platt}\ \emph {et~al.}(2012)\citenamefont {Platt},
  \citenamefont {Thomale}, \citenamefont {Honerkamp}, \citenamefont {Zhang},\
  and\ \citenamefont {Hanke}}]{platt2012}%
  \BibitemOpen
  \bibfield  {author} {\bibinfo {author} {\bibfnamefont {C.}~\bibnamefont
  {Platt}}, \bibinfo {author} {\bibfnamefont {R.}~\bibnamefont {Thomale}},
  \bibinfo {author} {\bibfnamefont {C.}~\bibnamefont {Honerkamp}}, \bibinfo
  {author} {\bibfnamefont {S.-C.}\ \bibnamefont {Zhang}}, \ and\ \bibinfo
  {author} {\bibfnamefont {W.}~\bibnamefont {Hanke}},\ }\href {\doibase
  10.1103/PhysRevB.85.180502} {\bibfield  {journal} {\bibinfo  {journal} {Phys.
  Rev. B}\ }\textbf {\bibinfo {volume} {85}},\ \bibinfo {pages} {180502}
  (\bibinfo {year} {2012})}\BibitemShut {NoStop}%
\bibitem [{\citenamefont {Trambly~de Laissardiere}\ \emph
  {et~al.}(2010)\citenamefont {Trambly~de Laissardiere}, \citenamefont
  {Mayou},\ and\ \citenamefont {Magaud}}]{trambly2010localization}%
  \BibitemOpen
  \bibfield  {author} {\bibinfo {author} {\bibfnamefont {G.}~\bibnamefont
  {Trambly~de Laissardiere}}, \bibinfo {author} {\bibfnamefont
  {D.}~\bibnamefont {Mayou}}, \ and\ \bibinfo {author} {\bibfnamefont
  {L.}~\bibnamefont {Magaud}},\ }\href@noop {} {\bibfield  {journal} {\bibinfo
  {journal} {Nano letters}\ }\textbf {\bibinfo {volume} {10}},\ \bibinfo
  {pages} {804} (\bibinfo {year} {2010})}\BibitemShut {NoStop}%
\bibitem [{\citenamefont {Stauber}\ and\ \citenamefont
  {Kohler}(2016)}]{stauber2016quasi}%
  \BibitemOpen
  \bibfield  {author} {\bibinfo {author} {\bibfnamefont {T.}~\bibnamefont
  {Stauber}}\ and\ \bibinfo {author} {\bibfnamefont {H.}~\bibnamefont
  {Kohler}},\ }\href@noop {} {\bibfield  {journal} {\bibinfo  {journal} {Nano
  letters}\ }\textbf {\bibinfo {volume} {16}},\ \bibinfo {pages} {6844}
  (\bibinfo {year} {2016})}\BibitemShut {NoStop}%
\end{thebibliography}%

\end{document}